\documentclass[twocolumn,showpacs,superscriptaddress,aps,prb]{revtex4-2}
\usepackage{bm}
\usepackage{amssymb,amsmath}
\usepackage{graphicx}
\usepackage{color}
\usepackage{marginnote}

\newcommand{\Xpp}{\chi_{\rho\rho}}
\newcommand{\Xiw}{\tilde\chi_{\rho\rho}}
\newcommand{\Xppargs}{\Xpp(\mathbf{x},\tau)}
\newcommand{\Xiwargs}{\Xiw(\mathbf{q},i\omega_m)}
\newcommand{\dynsuscargs}{\Xpp(\mathbf{q},\omega)}
\newcommand{\spec}{\chi_{\rho\rho}^{\prime\prime}}
\newcommand{\specargs}{\spec(\mathbf{q},\omega)}
\newcommand{\rhox}{\rho(\mathbf{x},\tau)}
\newcommand{\rhoxnot}{\rho(0,0)}

\newcommand{\spin}[2]{\mathbf{S}_{#1,#2}}

\begin{document}
\begin{abstract}
The amplitude (Higgs) mode near the two-dimensional superfluid-Mott glass quantum phase transition is studied. We map the Bose-Hubbard Hamiltonian of disordered interacting bosons onto an equivalent classical XY model in (2+1) dimensions and compute the scalar susceptibility of the order parameter amplitude via Monte Carlo simulation. Analytic continuation of the scalar susceptibilities from imaginary to real frequency to obtain the spectral densities is performed by a modified maximum entropy technique. Our results show that the introduction of disorder into the system leads to unconventional dynamical behavior of the Higgs mode that violates naive scaling, despite the underlying thermodynamics of the transition being of conventional power-law type. The computed spectral densities exhibit a broad, non-critical response for all energies, and a momentum-independent dispersion for long-wavelengths, indicating strong evidence for the localization of the Higgs mode for all dilutions. 
\end{abstract}

\title{Localization of the Higgs mode at the superfluid-Mott glass transition}
\author{Jack Crewse}
\author{Thomas Vojta}
\affiliation{Department of Physics, Missouri University of Science \& Technology, \\ Rolla, MO, 65409, USA}
\maketitle

\section{Introduction}
Zero-temperature phase transitions between quantum ground states of interacting many-body systems have become a central focus of modern condensed matter physics. The interest in these quantum phase transitions (QPTs) is justified by the rich physics that they exhibit, from unconventional thermodynamics and transport properties, to novel phases of matter. \cite{Sachdev, Sondhi_etal, Vojta2000, VojtaM} The effects of the inevitable disorder in condensed matter systems (impurities, defects, etc.) on these QPTs have also been intensely studied in the past two decades. Disorder leads to additional interesting physics, including infinite-randomness critical points \cite{Fisher92}, Griffiths singularities\cite{Griffiths, ThillHuse, RiegerYoung}, and smeared phase transitions\cite{HoyosVojta, Vojta2003} (for reviews see e.g. Refs. \onlinecite{Vojta2006, Vojta2010, Vojta2019}).

While much is understood about the thermodynamics of disordered QPTs, much less is known about the properties and dynamics of excitations near these critical points. Of particular interest are collective excitations in systems with spontaneously broken continous symmetry. A fundamental consequence of the breaking of the continuous symmetry of an $N$-component order parameter is the emergence of two distinct types of collective modes; the $(N-1)$ massless Goldstone modes -- fluctuations of the order parameter phase -- and a massive amplitude (Higgs) mode -- fluctuations of the order parameter amplitude.\cite{PekkerVarma,Burgess}  Prominent examples of condensed matter systems that exhibit this continuous symmetry breaking include Heisenberg and XY spin systems, superfluids, superconductors, and optical lattice bosons. Higgs excitations have also been observed experimentally in a number of these systems including: the superconductor NbSe$_2$ \cite{SooryakumarKlein}, the antiferromagnetic TiCuCl$_3$ \cite{Ruegg_etal}, and some incommensurate charge density wave compounds \cite{RenXuLupke,Pouget_etal}.

In Lorentz-invariant systems without disorder the Higgs mode is a sharp excitation in the ordered (broken symmetry) phase sufficiently close to the QPT, with a peak in the spectral density centered at the Higgs energy $\omega_H$. This energy softens as the critical point is approached. At zero wave vector, it obeys a power-law relationship controlled by the correlation length critical exponent $\omega_{H} \sim |r|^{\nu}$, where $r$ is the reduced distance from criticality. Higgs excitations in these clean systems have been widely studied.\cite{GazitPodolskyAuerbach, PodolskyAuerbachArovas} While the existence of a sharp Higgs peak in two-dimensions was initially in doubt, it was later proven by both analytic and numerical techniques. However, the fate of Higgs modes in the presence of disorder is much less understood.

In this article we therefore consider the effects of disorder on the Higgs mode excitation near the prototypical superfluid-Mott glass transition of disordered bosons. We model this transition using a particle-hole symmetric diluted quantum rotor model. This model is mapped onto an equivalent (2+1) dimensional classical XY model, which is then simulated via large-scale Monte Carlo methods. The imaginary (Matsubara) frequency scalar susceptibility of the order parameter is calculated. The associated spectral densities are found via analytic continuation of the Matsubara frequency data to the real-frequency axis via maximum entropy methods. 

Our results show that despite the critical behavior of the superfluid-Mott glass transition being of conventional power-law type, the Higgs mode shows unconventional dynamics that violates naive scaling. Specifically, the Higgs mode becomes strongly localized below the critical point for all dilutions, resulting in a broad non-critical response in the spectral densities arbitrarily close to the critical point. A short account of part of this work has already been published in Ref. \onlinecite{PuschmannCrewse_etal}.

The remainder of the article is organized as follows. In Section \ref{section:transition} we introduce the model Hamiltonian, the mapping to an equivalent classical model, and briefly discuss the thermodynamics of the corresponding superfluid-Mott glass transition. Section \ref{section:MonteCarlo} discusses the Monte Carlo simulations. Analytic continuation of the Matsubara frequency Monte Carlo data is detailed in Section \ref{section:Maxent} and the results discussed in Section \ref{section:Results}. We conclude and discuss experimental ramifications in Section \ref{section:Conclusion}.

\section{Superfluid-Mott glass transition}
\label{section:transition}

\begin{figure}
\includegraphics[width=\linewidth]{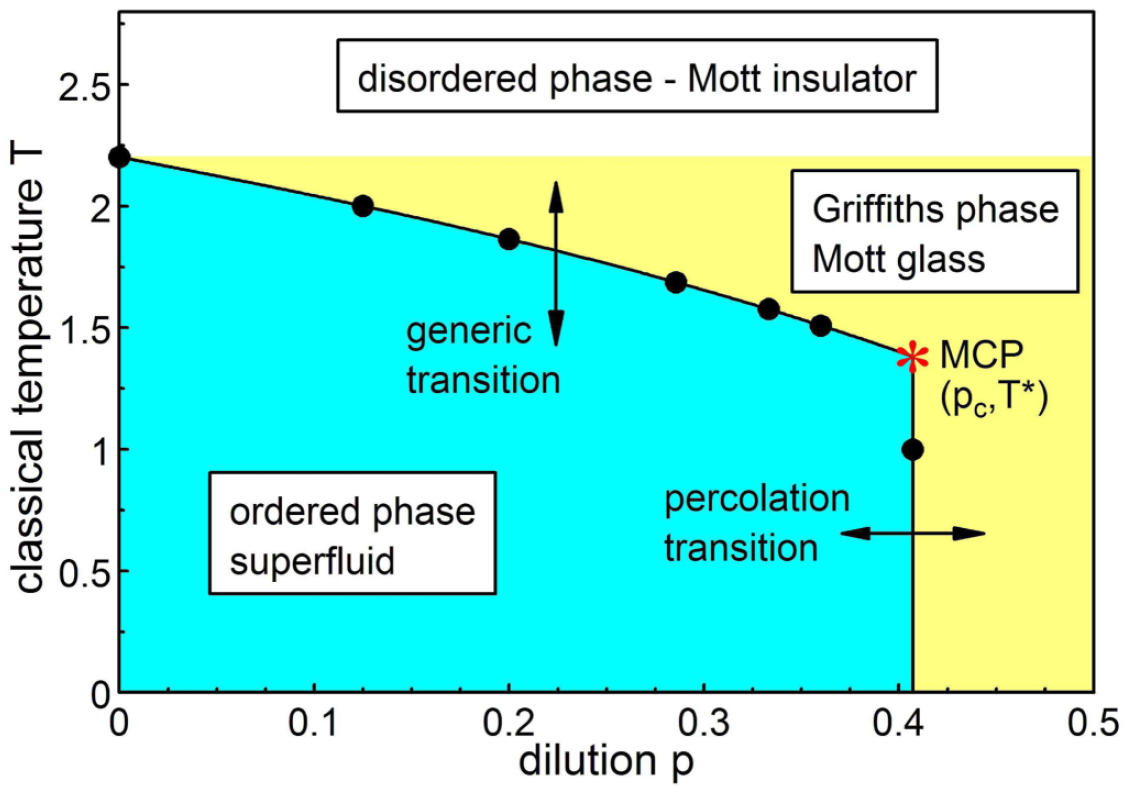}
\caption{Phase diagram of the classical (2+1)-dimensional XY model (\ref{classical-hamiltonian}) determined from Monte Carlo simulation.\cite{VojtaCrewse_etal} The emergence of the Mott glass phase is seen for arbitrarily small dilutions. Large dots mark the numerically calculated transitions, lines are spline fits that only serve as a visual guide. Here, we consider the Higgs mode for $p = 1/8, 1/5, 2/7, 1/3$ across these numerically determined generic transition points.}
\label{fig:phase-diagram}
\end{figure}

\begin{table}
	\begin{tabular}{|l|c|c|c|c|c|}
		\hline
		 & $z$ & $\beta/\nu$ & $\gamma/\nu$ & $\nu$ & $\eta$ \\ \hline
		Clean & $1$ & $0.5189(2)$ & $1.961(4)$ & $0.6717(1)$ & $0.0381(2)$ \\ \hline
		Diluted & $1.52(3)$ & $0.48(2)$ & $2.52(4)$ & $1.16(5)$ & $-0.52(4)$ \\ \hline
	\end{tabular}
	\caption{Critical exponents for the (2+1)d XY model. Clean exponents are from Ref. \onlinecite{Campostrini2006}. Disordered exponents are from Ref. \onlinecite{VojtaCrewse_etal}.}
	\label{table:critical-exponents}
\end{table}

We start from the Bose-Hubbard Hamiltonian describing bosons hopping between nearest-neighbor sites of a two-dimensional ($d = 2$) square-lattice of linear size $L$
\begin{equation}
\label{BH-hamiltonian}
H_{\text{BH}} = \frac{1}{2}\sum_{i} U_i (\hat{n}_i - \bar{n}_i)^2 - \sum_{\langle ij \rangle} J_{ij} (a^\dagger_i a_j + h.c.)
\end{equation}
where $a_i^{\dagger}$ and $a_i$ are bosonic creation and annihilation operators at a lattice site $i$ with $[a_i,a^\dagger_j] = \delta_{ij}$ and $\hat{n}_i = a^{\dagger}_i a_i$ as the number operator. Site-dependent interaction energy $U_i$, hopping amplitudes $J_{ij}$, and average filling $\bar{n}_i$ allow for a rich phase diagram. The phases and phase boundaries of this model have been well established via analytic methods.\cite{WeichmanMukhopadhyay} In the clean case of spatially uniform on-site interactions $U_i = U$, hopping amplitude $J_{ij} = J$, and average filling $\bar{n}_i = \bar{n}$ (excepting half-integer $\bar{n}$), the system exhibits a direct quantum phase transition between a superfluid ($J \gg U$) and a Mott insulating ($U \gg J$) ground state. Allowing spatially varied distributions (disorder) of $U_i$, $J_{ij}$ and $\bar{n}_i$ introduces a third, intermediate phase that separates the bulk superfluid and Mott insulating phases. The character of this intermediate phase is dependent on the qualitative nature of the distributions of $U_i$, $J_{ij}$, and $\bar{n}_i$. For generic disorder (realized, e.g., by random on-site potentials $\bar{n}_i$) the intermediate phase is the Bose glass, a compressible gapless insulator. If the disorder is such that the system is particle-hole symmetric (uniform integer $\bar{n}_i = \bar{n}$ and random $U_{i}$, $J_{ij}$), this intermediate phase instead becomes the \textit{incompressible} gapless Mott glass. 

We introduce disorder into the system with site-dilution by considering $U_i = U\epsilon_i$ and $J_{ij} = J\epsilon_i\epsilon_j$ where $U$ and $J$ are constants. The site-dilution is controlled then by the quenched random variables $\epsilon_i$ that take on the values $0$ (creates a vacancy) with probability $p$ and $1$ (creates an occupied lattice site) with probability $1-p$. If we consider the limit of large integer filling $\bar{n}_i=\bar{n}$, the Hamiltonian (\ref{BH-hamiltonian}) becomes equivalent to the Josephson junction (or quantum rotor) Hamiltonian
\begin{equation}
\label{JJ-hamiltonian}
H_{\text{JJ}} = \frac{U}{2}\sum_{i} \epsilon_{i}  \hat{n}_i^2 + J \sum_{\langle ij \rangle} \epsilon_i \epsilon_j \cos(\hat{\phi}_i - \hat{\phi}_j)
\end{equation}
where $\hat{n}_i$ now represents the fluctuations on top of the (uniform) filled background and $\hat{\phi}_i$ is the phase operator of a boson at site $i$. This model exhibits particle-hole symmetry for our site-dilution disorder and undergoes a QPT between the superfluid and Mott glass phases at a critical ratio $U/J$. 

To facilitate the study of the dynamics near the QPT via Monte Carlo simulation, we map the 2D quantum rotor Hamiltonian $H_{\text{JJ}}$ onto an equivalent classical model $H_C$ that is in the same universality class.\cite{Wallin_etal} This mapping yields a Hamiltonian with total dimensionality $D = d + 1 = 3$,
\begin{equation}
\label{classical-hamiltonian}
H_C = -J_s \sum_{\langle ij \rangle, \tau} \epsilon_i\epsilon_j\mathbf{S}_{i,\tau}\cdot\mathbf{S}_{j,\tau} - J_{\tau} \sum_{i,\tau} \epsilon_i\mathbf{S}_{i,\tau}\cdot\mathbf{S}_{i,\tau+1}
\end{equation}
with $\mathbf{S}_{i,\tau}$ as an $O(2)$ unit vector at space coordinate $i$ and imaginary-time coordinate $\tau$. The coupling constants are defined such that $\beta_CJ_s \sim 1/U$ and $\beta_CJ_{\tau} \sim J$ where $\beta_C = 1/T$ is the inverse temperature of the classical model. This mapping allows us to interpret the quantum model in two dimensions as a classical model at the inverse temperature $\beta_{C} = 1/T$ in three-dimensions. The temperature of the classical model is not the physical temperature of the quantum system (which is at absolute zero), but represents the ratio of the quantum coupling constants $U$ \& $J$ of the quantum system. Therefore, we can study the universal properties of the zero-temperature superfluid-Mott glass transition tuned by the ratio of couplings $U/J$, by tuning the classical temperature $T$ through the transition in the classical Hamiltonian $H_C$. For the remainder of this article, we will discuss the transition in $H_C$ in terms of the reduced distance from criticality $r = (T-T_c)/T_c$, for which the transition corresponds to $r \rightarrow 0$.

The thermodynamic critical behavior of $H_C$ falls into the 3D XY universality class for the undiluted case ($p=0$). The critical behavior in the presence of disorder was studied in Ref. \onlinecite{VojtaCrewse_etal}. It is of conventional finite-disorder type with a dynamical scaling characterized by a power law relation $\xi_\tau \sim \xi_s^z$ between the correlation lengths in space and imaginary-time. This is in contrast to many other disordered quantum phase transitions that feature ``infinite-randomness" critical points featuring activated dynamical scaling characterized by an exponentially growing relationship between the space and imaginary-time correlation lengths. 

The phase diagram of $H_C$ resulting from the simulations in Ref. \onlinecite{VojtaCrewse_etal} is presented in Figure \ref{fig:phase-diagram}.  The critical exponents for both the clean and diluted case can be found in table \ref{table:critical-exponents}. The numerically calculated critical exponents are used as inputs throughout the remainder of the article and careful consideration of their calculation, as well as the details of the phase diagram calculations can be found in our previous work.\cite{VojtaCrewse_etal}


\section{Monte Carlo simulation}
\label{section:MonteCarlo}
We study the Higgs mode by means of Monte Carlo simulation of the classical XY model Hamiltonian $H_C$. We consider a range of dilutions $p = 0, 1/8, 1/5, 2/7, 1/3$ below the lattice percolation threshold $p_c \approx 0.407253$. Dilutions higher than $p_c$ cause the lattice to form disconnected clusters and do not allow for any long range order formation. Both Metropolis\citep{Metropolis_etal} single-spin and Wolff\citep{Wolff} cluster algorithms are used throughout the simulation and one Monte Carlo sweep is defined by a Wolff cluster sweep plus a Metropolis sweep over the entire lattice.  A single Wolff sweep flips a number of clusters such that the total number of flipped spins is equal to the number of spins in the lattice. While the Wolff algorithm alone is sufficient in clean systems, highly dilute systems can exhibit small dangling clusters that the Metropolis algorithm can more effectively bring to equilibrium. 

\begin{figure}
\includegraphics[width=\linewidth]{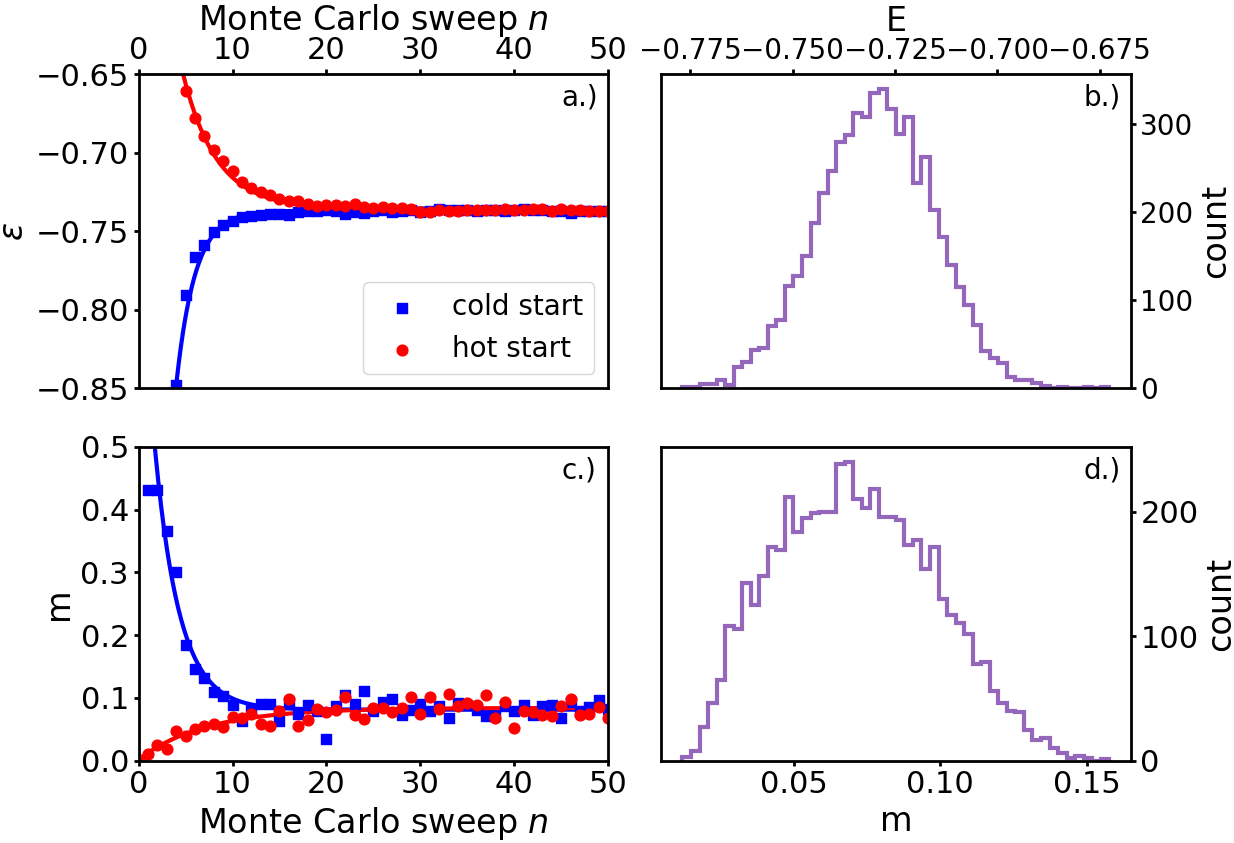}
\caption{Comparison of `hot' (randomly aligned spins) and `cold' (aligned spins) start equilibration times for a.) energy per particle $\varepsilon = E/V$ (where $V$ is the number of spins), and c.) order parameter $m$ for a highly-dilute system ($p = 1/3$) at criticality $T = T_c = 1.5735$. Calculated from a single disorder realization of size $L=100$ and $L_{\tau}=452$. Fits of the energy and order parameter data (solid lines) to an exponential form yields equilibration times $\tau_{eq} \sim 3-7$. Histograms of the energy per particle and order parameter for $6000$ disorder realizations of the same system are shown in b.) and d.), respectively. Each disorder realization was averaged over $1000$ Monte Carlo sweeps to obtain accurate estimates.}
\label{fig:equilibration}
\end{figure}

We estimate equilibration times by directly analyzing the evolution of the energy per particle $E/V$ and order parameter $m = \frac{1}{V}\sum_i \mathbf{S}_i$ as a function of Monte Carlo sweep $n$ (where $V$ the number of occupied lattice sites). Figures \ref{fig:equilibration}a and \ref{fig:equilibration}c shows this evolution for a case where equilibration is expected to take the longest -- a large, highly-dilute system right at criticality. The energy and order parameter reach equilibrium values for both a `hot start' (all spins randomly oriented) and a `cold start' (all spins aligned) after only $n \approx 30$ Monte Carlo sweeps. Fitting the energy data to $E_n = E_{av} + a\exp(-n/\tau_{eq})$ (and analogously for order parameter) results in equilibration times not exceeding $\tau_{eq} = 8$. We choose a number of equilibration sweeps many times larger than any measured equilibration times $N_{eq} = 100$ to ensure measurements are taken on properly equilibrated systems for even the most extreme disorder realizations. 

Distributions of key thermodynamic observables has also been considered to ascertain the significance of rare events. Histograms of the energy per particle and order parameter are presented in Figure \ref{fig:equilibration}b and \ref{fig:equilibration}d. While the distributions are moderately broad, they feature no long tails. This is in agreement with what is expected for a finite-disorder fixed point\cite{VojtaHoyos, VojtaSchmalian}, for which the conventional power-law type critical behavior is the superfluid-Mott glass transition implies.\cite{VojtaCrewse_etal} 

Due to the large computational effort required to simulate disordered systems, we carefully consider the balance of measurement steps $N_M$ and disorder realizations (samples) $N_S$.\cite{Ballesteros_etal, VojtaSknepnek} The final variance $\sigma^2$ of a given observable after both the thermodynamic (Monte Carlo) and disorder averaging can be estimated as
\begin{equation}
\sigma^2 \approx (\sigma_S^2 + \sigma_M^2/N_M)/N_S
\end{equation}
where $\sigma_S^2$ is the disorder-induced variance and $\sigma_M^2$ is the variance of single measurement for a given disorder realization. Computational effort is roughly proportional to $(N_M + N_{eq})N_S$, thus we can achieve best performance with a reasonably small $\sigma^2$ by considering a large number of disorder realizations with a relatively small number of measurement steps. In our simulations we choose $N_M = 500$ with a number of disorder realizations $N_S = 5000-10000$ (dependent on system sizes). 

The small number of measurement steps comes at the cost of introducing biases to traditional estimators of the required correlation functions. Without the need for disorder averaging the bias decays much faster ($\sim N_M^{-1}$) than the statistical error ($\sim N_M^{-1/2}$) and can be neglected for long Monte Carlo runs. Averaging short runs over a large number of disorder realizations suppresses the decay of the statistical error by another factor of $N_S^{-1/2}$, thus the bias may become commensurate to the statistical error and must be considered. To eliminate these biases we utilize improved estimators as discussed e.g. in Ref. \onlinecite{Zhu_etal}.

As the introduction of quenched disorder breaks the isotropy between the space and imaginary-time dimensions in the Hamiltonian (\ref{classical-hamiltonian}), the standard finite-size scaling techniques to calculate critical exponents breaks down in the disordered case. There are two characteristic length scales we must consider in the simulations: the spatial correlation length $\xi_s$ and the correlation length in imaginary-time $\xi_{\tau}$. Correspondingly, the system sizes in the spatial dimensions $L$ and the imaginary-time dimension $L_{\tau}$ are independent parameters. Anisotropic two-parameter finite-size scaling needs to be used to find the ``optimal" aspect ratios $L_{\tau}/L^z$ (equivalently determining the dynamical exponent $z$), by considering system sizes that maximize the Binder cumulant at the quantum critical point (QCP). We utilize the results for the ``optimal shapes" obtained in our previous simulations of the thermodynamic critical behavior.\cite{VojtaCrewse_etal} Further technical details can be found in Ref. \onlinecite{VojtaCrewse_etal}, as well as other works on the critical behavior of Ising spin glasses. \cite{GuoBhattHuse}

To suppress any finite-size effects, we consider only the largest system sizes accessible within our computational limits. We consider spatial sizes up to $L = 100$ and imaginary-time sizes up to $L_{\tau}=452$ for diluted systems. These system sizes exceed the correlation lengths and times of the excitations we examine. For example, the smallest Higgs energy calculated for the clean case is $\omega_H \approx 0.21$ giving a characteristic time of $2\pi/\omega_H \approx 30$, much smaller than any of the imaginary-time system sizes used. Finite-size effects in the disordered case are of even lesser concern as our results suggest that the Higgs mode localizes, and the energy of the Higgs spectral peak remains microscopic  (see Fig. \ref{fig:dirty_specs}).


\section{Scalar susceptibility and spectral densities}
\label{section:Scalar_spectral}
The amplitude mode is a collective excitation of the order parameter \textit{magnitude}. The local degrees of freedom of the system defined by (\ref{classical-hamiltonian}) are of fixed magnitude $|\mathbf{S}_{i,\tau}| = 1$, so we must define a local order parameter that can fluctuate. We define our order parameter by considering a course-graining of the local degrees of freedom. This is calculated as the vector sum of the $\mathbf{S}_{i, \tau}$ at the site $i$ with its nearest (spatial) neighbors.\footnote{We have also considered an alternative definition of the order parameter which includes the next-nearest-neighbors. Qualitative behavior of the Higgs mode is unaffected in both the clean and disordered systems.} It's magnitude reads
\begin{equation}
\rho(\mathbf{x}_i,\tau) = \frac{1}{5}\bigg|\epsilon_i\spin{i}{\tau} + \sum_j^{n.n.}\epsilon_j\spin{j}{\tau}\bigg|.
\end{equation}
Information about the Higgs mode is contained in the imaginary-time scalar susceptibility of the local order parameter magnitude $\rhox$
\begin{equation}
\label{Xpp}
\Xppargs = \langle\rhox\rhoxnot\rangle - \langle\rhox\rangle\langle\rhoxnot\rangle
\end{equation}
and it's Fourier transform $\Xiwargs = \int d\mathbf{x}d\tau e^{-i\mathbf{q} \cdot \mathbf{x} - i\omega_m \tau}\Xppargs$ in terms of Matsubara frequencies $\omega_m  = 2\pi m /\beta$ and wave vector $\mathbf{q}$. The real-frequency dynamic susceptibility is obtained via analytic continuation
\begin{equation}
\dynsuscargs = \Xiw(\mathbf{q},i\omega_m \rightarrow \omega + i0^+).
\end{equation}
The spectral density, which is related to many experimental probes, is then proportional to the imaginary part of the dynamic susceptibility
\begin{equation}
\specargs = \operatorname{Im} \dynsuscargs.
\end{equation}

A scaling form for the real-frequency susceptibility at the \textit{clean} superfluid-Mott insulator transition has been derived by Podolsky and Sachdev.\cite{PodolskySachdev} This can be generalized to include the quenched disorder and an appropriate dynamical exponent for the \textit{diluted} transition we are interested in.  We start from a $d$-dimensional, quantum field theory for an $N$-component order parameter $\psi$ defined by the action
\begin{equation}
\label{action}
S = \int d^dxd\tau [(\partial_\mathbf{x}\psi)^2 + (\partial_{\tau}\psi)^2 + (r + \delta r(\mathbf{x}))\psi^2 + u\psi^4]
\end{equation}
where $r$ is the reduced distance from criticality, $\delta r(\mathbf{x})$ represents a quenched random-mass disorder and $u$ is the quartic interaction strength. For the parameters of our system, $d=2$ and $N=2$, the action (\ref{action}) is a coarse-grained, long-wavelength approximation of the quantum rotor model (\ref{JJ-hamiltonian}) and exhibits a QPT in the same universality class.

The free energy is given as
\begin{equation}
f = -\frac{1}{\beta V} \ln Z = -\frac{1}{\beta V} \ln \int D[\psi] e^{-S}.
\end{equation} 
We then notice that with two derivatives of this free energy with respect to the distance from criticality, we arrive at the expression
\begin{align*}
\frac{d^2f}{dr^2} =& \frac{1}{\beta V} \int d^dx d\tau \int d^dx^\prime d\tau^\prime  \\
\times &  [\langle\psi^2(\mathbf{x},\tau) \psi^2(\mathbf{x}^\prime , \tau^\prime)\rangle - \langle\psi^2(\mathbf{x},\tau)\rangle \langle \psi^2(\mathbf{x}^\prime , \tau^\prime)\rangle] 
\end{align*}
which is the exact expression for the $\mathbf{q} = 0$, $\omega_m = 0$ Fourier components of the scalar susceptibility of the order parameter $\Xpp$. More precisely, this yields the susceptibility of the square of the order parameter amplitude, however as the order parameter magnitude is non-zero at criticality, the scaling behavior of both correlation functions is the same. 
The singular part of the free energy fulfills the homogeneity relationship
\begin{equation}
\label{free-energy}
f(r) = b^{-(d+z)}f(rb^{1/\nu})
\end{equation}
with $b$ as an arbitrary scale factor. From the argument above, taking two derivatives of the free energy (\ref{free-energy}) with respect to $r$ gives the scaling behavior of the scalar susceptibility, thus implying the scaling form 
\begin{equation}
\Xpp(r,\mathbf{q},\omega) = b^{-(d+z) + 2/\nu}\Xpp(rb^{1/\nu}, \mathbf{q}b,\omega b^z)
\end{equation}
from which we identify the scale dimension of $\Xpp$ as $-(d+z) + 2/\nu$. Setting $b=r^{-\nu}$ we arrive at the scaling form
\begin{equation}
\label{naive-scaling-r}
\Xpp(r,\mathbf{q},\omega) = r^{(d+z)\nu - 2} X(\mathbf{q}r^{-\nu},\omega r^{-z\nu})
\end{equation}
or equivalently, with $r \sim \omega^{1/z\nu}$
\begin{equation}
\label{naive-scaling}
\Xpp(r,\mathbf{q},\omega) = \omega^{[(d+z)\nu - 2]/(\nu z)} Y(\mathbf{q}r^{-\nu},\omega r^{-z\nu})
\end{equation}
where $X$ and $Y$ are scaling functions, and $z$ is the dynamical critical exponent. If we set the dynamical exponent to the clean value $z=1$ in two-dimensions $d=2$ in equation (\ref{naive-scaling-r}), we arrive at the scaling form derived by Podolsky \& Sachdev for the clean superfluid-Mott insulator transition
\begin{equation}
\label{clean-scaling}
\dynsuscargs = r^{3\nu - 2} X(\mathbf{q}r^{-\nu}, \omega r^{-\nu}).
\end{equation}

Considering the critical exponents calculated for the two-dimensional superfluid-Mott glass transition, the scaling form (\ref{naive-scaling}) makes some interesting predictions about the fate of the Higgs mode in the diluted case. For our case, using the critical exponents calculated for the diluted transition, $z = 1.52$ and $\nu  = 1.16$ (see table \ref{table:critical-exponents}), we see that we have
\begin{equation}
[(d+z)\nu - 2]/(\nu z) \approx 1.18 > 0.
\end{equation}
This positive scaling dimension suggests that the amplitude of the singular part of the scalar susceptibility becomes strongly suppressed as the critical point is approached. Thus, the introduction of disorder may destroy a sharp, well-defined Higgs mode excitation near the QCP. 

This argument can be extended to any quantum system with random mass disorder. The condition for a strongly suppressed Higgs peak near the QCP $[(d+z)\nu - 2]/(\nu z) > 1$ is equivalent to the condition $d\nu > 2$. It is well known that general disordered systems must satisfy the inequality $d\nu \ge 2$.\cite{Chayes_etal} Thus it is guaranteed that we have a scaling dimension $[(d+z)\nu - 2]/(\nu z) > 1$ for the scalar susceptibility, strongly suppressing the singular part of $\Xpp$ in the excitation spectra as the QCP is approached.

\newpage

\section{Maximum Entropy Methods}
\label{section:Maxent}
The Monte Carlo simulations output the scalar susceptibility $\Xiwargs$ as a function of Matsubara frequency $\omega_m = 2\pi m/\beta$. The spectral densities $\specargs$ we are interested in are related to the scalar susceptibilities by the Kramers-Kronig relationship
\begin{equation}
\label{maxent-integral}
\Xiwargs = \frac{1}{\pi} \int_0^\infty d\omega \specargs \frac{2\omega}{\omega_m^2 + \omega^2}.
\end{equation}
In principle, one could invert this relationship to extract the spectral densities from the computed scalar susceptibility directly. Unfortunately, this inversion is ill-conditioned and the inevitable noise of Monte Carlo data only exaggerates the problem (small errors in the input data can create large features in the spectral density). 

To overcome this issue, we use a modified maximum entropy (MaxEnt) method.\citep{JarrellGubernatis, Gazit_etal} The method utilizes Bayesian inference to transform the integral inversion problem (\ref{maxent-integral}) into finding the most probable spectral density given the input quantum Monte Carlo (QMC) data. This reduces the problem to minimizing a cost function
\begin{equation}
  \label{maxentQ}
  Q = \Delta - \alpha S.
\end{equation}
The first term in $Q$,
\begin{equation}
\label{maxentD}
\Delta = (\Xiw - K\spec)^T\Sigma^{-1}(\Xiw - K\spec)
\end{equation}
serves as a measure of how well the fitted spectral density $\spec$ reproduces the input data $\Xiw$. Here, $K$ is a discretized version of the integration kernel $K(\omega,\omega_m) = 2\omega/(\omega_m^2 + \omega^2)$ and $\Sigma_{mn} = \langle \Xiw(i\omega_m) \Xiw(i\omega_n) \rangle - \langle \Xiw(i\omega_m) \rangle \langle\Xiw(i\omega_n) \rangle$ is the covariance matrix of the scalar susceptibility data. The second term is an entropy of the spectral density
\begin{equation}
\label{maxentS}
S = - \sum_\omega\spec(\omega)\ln\spec(\omega)
\end{equation}
that serves to regularize the inversion process, preventing over-fitting of Monte Carlo noise. This regularization is achieved because large entropy values favor a smooth spectral density, thus punishing over-fitting of the unphysical noise in the minimization of $Q$.

\begin{figure}
\includegraphics[width=\linewidth]{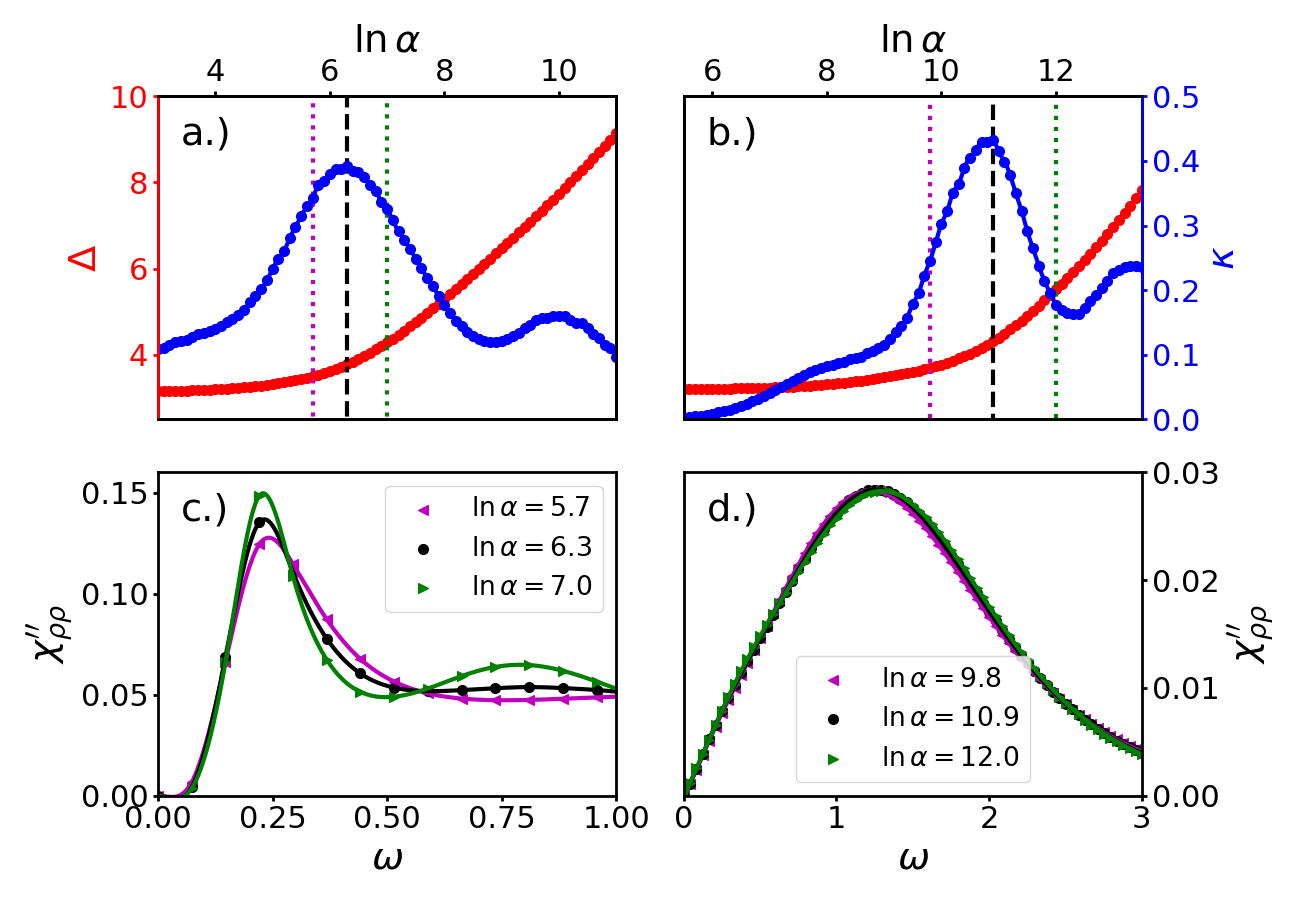}
\caption{Maximum entropy method and it's sensitivity to variation with respect to the fit parameter $\alpha$. a.) the error sum $\Delta$ and it's curvature $d^2\Delta / d(\ln \alpha)^2$ vs. $\ln\alpha$ for a clean system of size $L=L_{\tau}=128$ a distance $r=-0.01$ from criticality. Dashed lines represent our chosen optimal fit parameter $\ln\alpha = 6.3$, dotted lines represent the range of variation we consider in error estimations. b.) Similar data for a diluted ($p=1/3$) system of size $L=100$, $L_{\tau}=452$ with analogous parameters. c.) Spectral densities for the clean system at the values of $\ln\alpha$ indicated in a.). d.) Spectral densities for diluted ($p=1/3$) system for values of $\ln\alpha$ indicated in b.). }
\label{fig:maxent-process}
\end{figure}

This leaves an additional free parameter $\alpha$ that controls the relative weight between the goodness-of-fit term $\Delta$ and the entropy term in $Q$. There are a number of choices in the literature concerning the determination of the value of $\alpha$ for a given fit. In our calculations we choose the value of $\alpha$ by a version of the L-curve method (see Figs. \ref{fig:maxent-process}a-b) which maximizes the curvature $\kappa = d^2\Delta / d(\ln \alpha)^2$.\cite{BergeronTremblay,HansenOleary} This maximum marks a crossover from the fitting of information to the fitting of noise. Additional methods of determining the optimal fit parameter choose $\alpha$ such that $\Delta$ is roughly equal to the number of independent Matsubara frequencies $\omega_m$ being fit. In our simulations, we use this condition as a check for the suitability of the optimal alpha found by maximizing the curvature.

\subsection{Discrete time-step modifications of the Maxent method}
The integral relationship (\ref{maxent-integral}) we seek to invert in the maximum entropy method assumes continuous time or, equivalently, an infinite set of Matsubara frequencies. However, our quantum Monte Carlo method works in discrete imaginary-time. Some previous implementations of this method have used spline interpolation of the discrete time Monte Carlo data to best approximate a continuous input $\Xiw$.\cite{BergeronTremblay} While this choice allows simple numerical integration when calculating (\ref{maxentD}), the interpolation method may introduce additional uncertainties that are not accounted for.
In our calculations we take a different approach, instead modifying the integral kernel in (\ref{maxent-integral}) to account for both the discrete nature of the data as well as the periodic boundary conditions imposed on the system in the simulations. \cite{Gazit_etal}

The QMC data from simulation are time discrete with imaginary-time values $\tau_k = k\Delta \tau$ with $k \in \{0,1,...,N-1\}$ where $N = \beta/\Delta \tau$. For a given general imaginary-time Green's function $G(\tau_k)$ the discrete Fourier transform in terms of Matsubara frequencies $\omega_m = 2\pi m/\beta = 2\pi m/\Delta \tau N$ is given by
\begin{equation}
\label{fourier_greens}
G(i\omega_m)  = \sum_{k=0}^{N-1} e^{i2\pi \omega_m \tau_k} G(\tau_k) \Delta \tau = \Delta \tau \sum_{k=0}^{N-1} e^{i2\pi m k/N} G(\tau_k) 
\end{equation}
The spectral (Lehmann) representation of an arbitrary imaginary-time Green's function involving operators $A$ and $B$ is given by
\begin{equation}
\label{greens_lehmann}
G(\tau_k) = - \langle A(\tau_k)B(0)\rangle = \frac{1}{Z} \sum_{l,m} A_{lm}B_{ml} e^{\tau_k (E_l - E_m)} e^{-\beta E_l}
\end{equation}
where $Z = \text{Tr} (e^{-\beta H})$ is the partition function of the system and $E_n$ the energy of eigenstate $|n\rangle$. If we insert this into (\ref{fourier_greens}), and carry out the sum over $k$, we arrive at
\begin{equation}
\label{greens_discrete}
G(i\omega_m) = \frac{\Delta \tau}{Z}\sum_{l,m} A_{lm}B_{ml} 
								\bigg[
								\frac{e^{-\beta E_l} - e^{-\beta E_m}}
								         {e^{\Delta \tau (i\omega_m + E_l - E_m) - 1}}
								\bigg]
\end{equation}
which allows the identification of the spectral density
\begin{equation}
\label{spectral_lehmann}
\spec(\omega) = \frac{1}{Z}\sum_{l,m} A_{lm}B_{ml} 
								[e^{-\beta E_l} - e^{-\beta E_m}] \delta (\omega - E_m - E_l).
\end{equation}
Equation (\ref{spectral_lehmann}) is the same result as one would find in the continuous time case. This justifies expressing the discrete Matsubara frequency Green's function in terms of the continuous time spectral density such that we have
\begin{equation}
\label{greens_discrete_integral}
G(i\omega_m) = \int_{-\infty}^{\infty} d \omega \spec(\omega) \frac{\Delta \tau}{e^{\Delta \tau (i\omega_m - \omega) - 1}}.
\end{equation}

We now take advantage of expected properties of the spectral density $\spec(\omega)$. For bosonic operators we have $\spec(-\omega) = -\spec(\omega)$. This allows us to split the integration for (\ref{greens_discrete_integral}) into
\begin{equation}
G(i\omega_m) = \int_{0}^{\infty} d \omega \spec(\omega) \bigg[ \frac{\Delta \tau}{e^{\Delta \tau (i\omega_m - \omega) - 1}} -  \frac{\Delta \tau}{e^{\Delta \tau (i\omega_m + \omega) - 1}} \bigg].
\end{equation}
which simplifies to the form used in the MaxEnt procedure
\begin{equation}
G(i\omega_m) = \int_{0}^{\infty} d\omega \spec(\omega) \bigg[ \frac{\Delta \tau \sinh(\Delta \tau \omega)}{\cos(\Delta \tau \omega_m) - \cosh(\Delta \tau \omega)}\bigg]
\end{equation}

We can verify that this simplifies to the appropriate continuous time case for $\Delta \tau \rightarrow 0$, by expanding the trigonometric functions for $\Delta \tau \ll 0$
\begin{equation}
G(i\omega_m) \xrightarrow{\Delta \tau \rightarrow 0} - \int_{0}^{\infty} d\omega \spec(\omega) \frac{2\omega}{\omega_m^2 + \omega^2}.
\end{equation}
The convergence to the continuous time case also holds along each step of the derivation. 

Applying these modifications to (\ref{maxent-integral}) one arrives at the relationship that is used in our MaxEnt procedure
\begin{equation}
\Xiwargs = \frac{1}{\pi} \int_0^{\infty} d\omega \specargs \frac{\Delta \tau \sinh(\Delta \tau \omega)}{\cos(\Delta \tau \omega_m)-\cosh(\Delta \tau \omega)}
\end{equation}
It is straight forward to confirm that if one takes $\Delta \tau \rightarrow 0$, we arrive back at the continuous time form (\ref{maxent-integral}). 

\subsection{Maximum Entropy error}
\label{subsection:MaxEntError}
The maximum entropy technique is notoriously sensitive to both the noise of the QMC data and the choice of weight parameter $\alpha$. To understand the extent to which the output spectral densities are sensitive to these two sources of error, we have utilized two methods to make estimates of the total error introduced in this numeric analytic continuation. 

First, we utilize an ensemble method to estimate the sensitivity of the method to the input QMC error bars $\sigma_i$. In addition to the analytic continuation of the output QMC data, we generate a set of synthetic data by adding random variables drawn from a Gaussian distribution with width equal to the error bar $\sigma_i$ of the calculated data point. A separate Maxent procedure is performed for each of these synthetic data sets. A statistical error is then calculated from an ensemble average and variance of the resulting set of spectral densities. This gives an estimate of the statistical error introduced in the maximum entropy process from the QMC data. These error bars will be presented in each figure in section \ref{section:Results}, however the magnitude of the error is such that the error bars are smaller than the symbol sizes.  

Second, we consider how variation of the weight parameter $\alpha$ about the neighborhood of the optimal value $\alpha^*$ affects the resulting spectral density. Taking values $\pm 5\%$ of $\ln\alpha^*$, we find that the spectral density peak positions are only weakly varied in the frequency axis, with a variation of only a few percent of their peak energies at $\alpha^*$ for small wave vector $q$. For larger $q$, the spectral densities become much more broad and have significantly smaller amplitude. As the peak positions of a broad maximum is less well-defined, this leads to a larger variation in the apparent peak positions for short wavelengths. This can be seen as the error bars presented with the dispersion data in Figure \ref{fig:dispersion}. The peak amplitudes are significantly more sensitive to the exact value of $\alpha^*$ in the clean case with variations of the peak amplitudes $\approx 10\%$ (Fig. \ref{fig:maxent-process}c). In the diluted case the broader peaks and significantly smaller amplitudes suppress the peak amplitude variations (Fig. \ref{fig:maxent-process}d). 

Lastly, we consider the effects introduced by changing the number of fitted Matsubara frequency data points. This becomes important in the case of large $q$, as the peak frequencies begin to increase and have their features broadened. For small $q$, the main features in the spectral density are at low frequency and, thus only a few percent variation is observed in the spectral densities when the number of $\omega_m$ included is changed, with most of the difference seen in the tails leaving the peaks relatively unaffected. For large $q$, an increasing number of fitted Matsubara frequencies is required to capture the main features of the broader, high-frequency features of the spectral densities. We therefore utilize all available Matsubara frequencies for fitting when considering the dispersion for the full range of $q$. 

\section{Results: Higgs mode localization}
\label{section:Results}
\begin{figure}
\includegraphics[width=\linewidth]{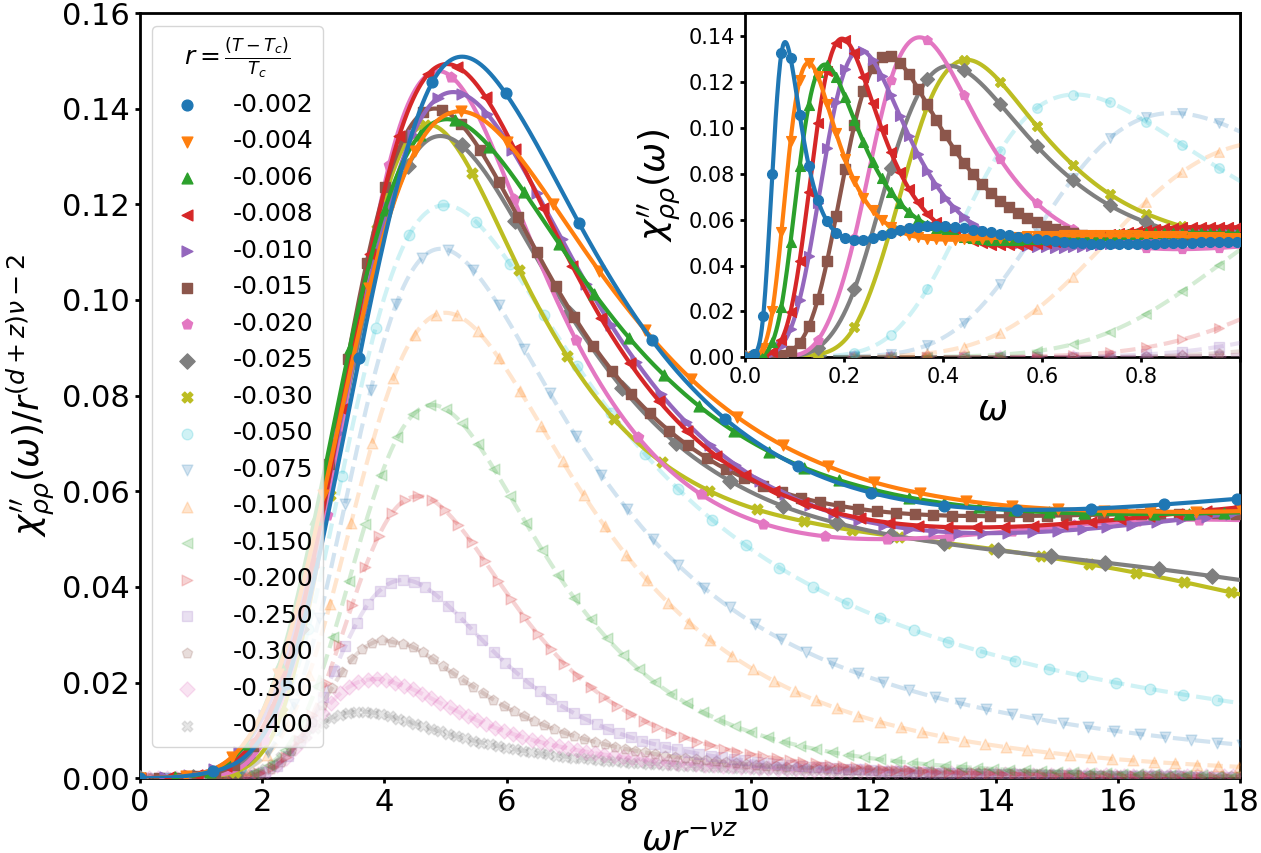}
\caption{Spectral densities in the clean ($p=0$) case for $\mathbf{q} = 0$ at different distances from criticality $r < 0$. Calculated from a system of size $L=L_{\tau}=128$ and averaged over $10,000$ samples. Main panel: Spectral densities scaled according to the expected scaling form (\ref{clean-scaling}). The clean spectral densities scale as expected to within the error bars of the maximum entropy techniques. Inset: Raw data of spectral densities showing the softening of $\omega_H$ as the QCP is approached. Spectral densities outside the scaling window ($r < -0.30$) are indicated by faint/dotted line plots.}
\label{fig:clean_specs}
\end{figure}

For the clean ($p=0$) system, the amplitude mode is seen as a well-defined, soft-gapped excitation with a sharp peak in the spectral density $\chi_{\rho\rho}^{\prime\prime}$ centered at the Higgs energy $\omega_H$. The calculated zero-wavenumber spectral densities are shown in Figure \ref{fig:clean_specs} for a range of distances from criticality $r = (T-T_c)/T_c$. The expected scaling behavior (\ref{clean-scaling}) is seen to be satisfied by the collapse of the spectral densities with respect to both Higgs energy $\omega_H$ and amplitude for $r \ge -0.030$. The remaining variation between the curves in this $r$-range is within the errors introduced by the maximum entropy method. Beyond $r < -0.030$, both Higgs energy and amplitudes begin to violate scaling. This can be attributed to being outside of the critical window where scaling is not expected to be satisfied. These results are in agreement with previous studies of the clean case Higgs mode.\cite{Gazit_etal}

\begin{figure}
\includegraphics[width=\linewidth]{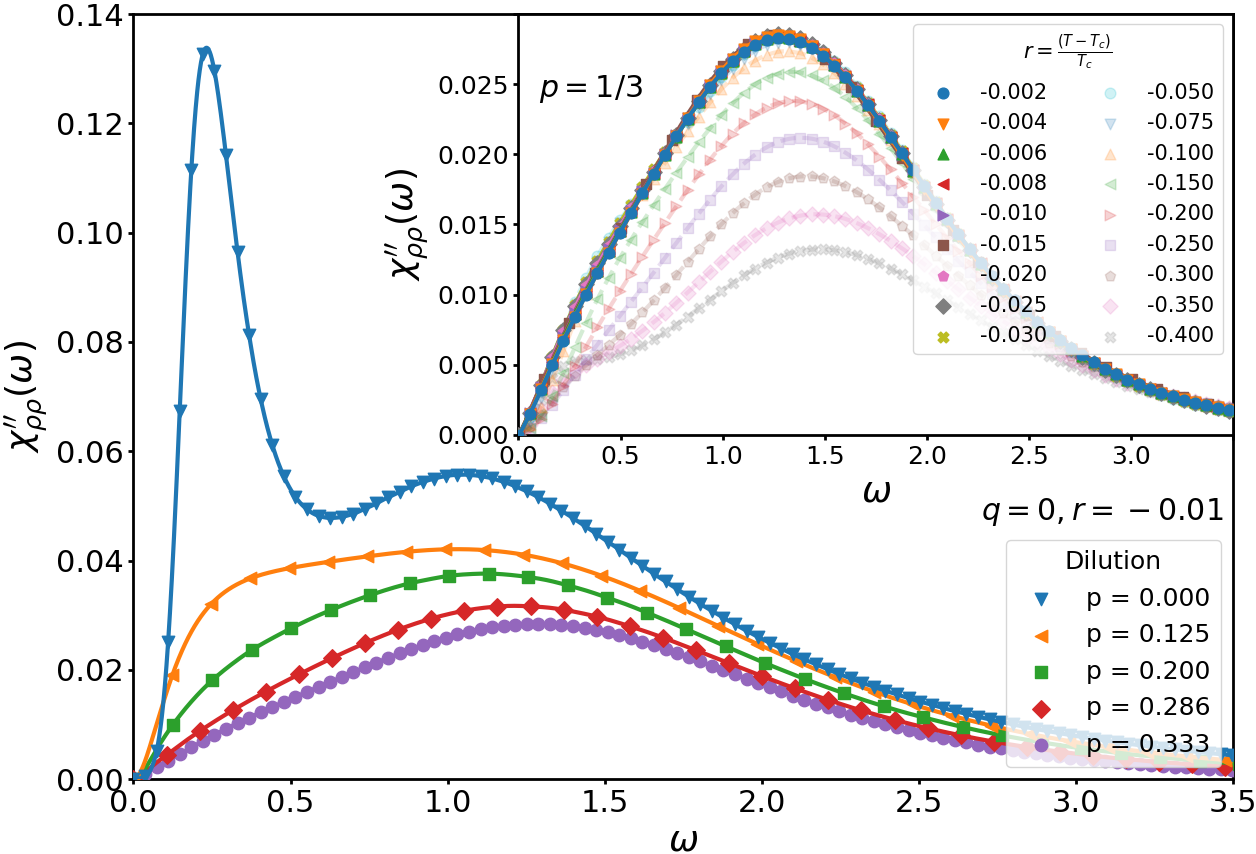}
\caption{Main panel: Zero-wavenumber $\mathbf{q} = 0$	 spectral densities at a fixed distance from criticality $r = -0.01$ for each of the dilutions considered. For each dilution we simulate a system size $L=100$ (with $L_{\tau} = 175, 241, 358, 452$ corresponding to each dilutions ``optimal'' system size fixed by the dynamical exponent $z$) averaged over $10,000$ disorder samples. The prominent Higgs peak seen in the clean ($p=0$) case is not observed even in the lowest dilutions considered ($p=1/8$). Inset: Spectral densities of a highly-dilute system ($p=1/3$) as a function of $r$. Within the scaling region $|r| < 0.03$, the spectral densities show no dependence on distance from criticality. }
\label{fig:dirty_specs}
\end{figure}

A much different behavior is seen in the spectral densities as soon as disorder is introduced to the system. Our calculated zero-wavenumber spectral densities can be seen in Figure \ref{fig:dirty_specs}. The main panel compares $\spec$ for several dilutions at a fixed distance from criticality $r=-0.01$. A broad, non-critical response in the spectral densities is seen for all dilutions below the percolation threshold, with no sharp Higgs peak present. Even for the smallest dilution considered ($p=1/8$), the Higgs peak is strongly suppressed, with the main contribution to the spectral weight being at high-frequencies. As dilution is increased the high-frequency contribution quickly dominates the spectral weight as can be seen in the main panel of Figure \ref{fig:dirty_specs}. For larger dilutions $\spec$ becomes almost dilution independent. Even more interesting is the dependence of $\spec$ on the distance from criticality. In contrast to the clean case, only weak variation of the spectral weight is observed as distance from criticality is adjusted for the smallest dilutions. For higher dilutions, this variation within the critical scaling region is effectively non-existent as can be seen for $p=1/3$ in the inset of Figure \ref{fig:dirty_specs}. This response of the diluted system clearly violates the naive scaling form (\ref{clean-scaling}) further indicating that the spectral densities must be dominated by some non-critical contribution that does not feature a sharp Higgs peak. 

\begin{figure}
\includegraphics[width=\linewidth]{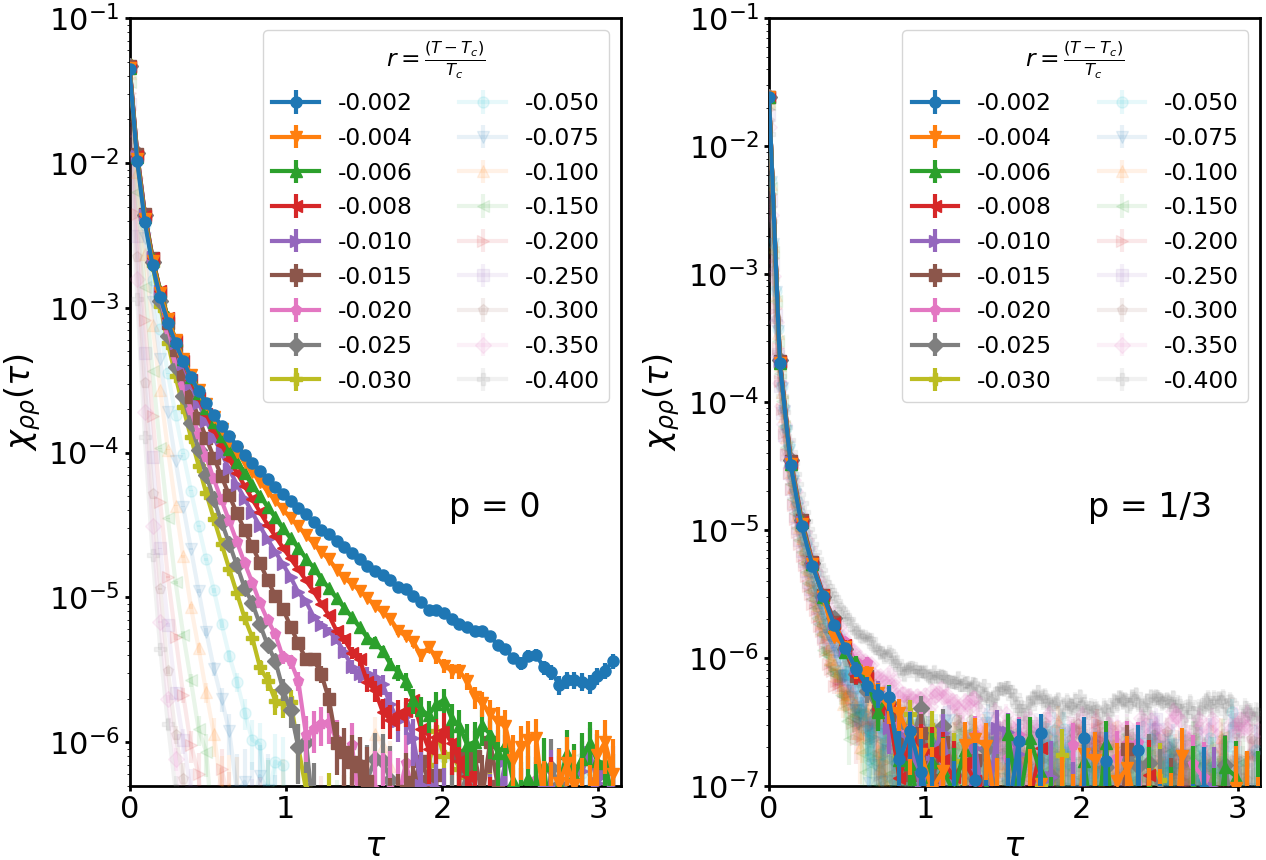}
\caption{Comparison of scalar correlation functions in imaginary-time for a clean system (left) and a highly-dilute system (right). Approximately exponential decay of $\Xpp(\tau)$ for long times implies a well-defined single frequency peak in the associated spectral density. Parameters are as in Fig. \ref{fig:dirty_specs}.}
\label{fig:it-correlations}
\end{figure}

This striking difference between the clean and diluted spectral densities is already evident in the imaginary-time scalar correlation functions themselves where the potential instabilities of the MaxEnt procedure are of no concern. Figure \ref{fig:it-correlations} shows the correlation functions $\Xpp(\tau)$ with respect to imaginary-time for both the clean and highly-dilute case. In the clean case, the approximately exponential decay of the correlations for large $\tau$ implies a well-defined, single-frequency peak in the associated spectral densities.\footnote{The long-time decay is not purely exponential as the clean spectral density is known to feature a soft gap at $\omega = 0 $ rather than a hard gap.} The softening of $\omega_H$ as the critical point is approached is also easily observed, with increasing decay times closer to criticality. In contrast, the diluted case shows a much faster, non-exponential decay of the correlation function, implying a broad frequency response in the spectral densities. Additionally, the absence of any dependence on distance from criticality is observed, with correlation functions within the region $|r| \le 0.03$ effectively identical within error bars. These two features of the diluted correlations functions are in agreement with the response seen in the spectral densities, verifying that the unconventional nature of the Higgs mode in the diluted system is evident even before the uncertainties of the MaxEnt process. 

\begin{figure}
\includegraphics[width=\linewidth]{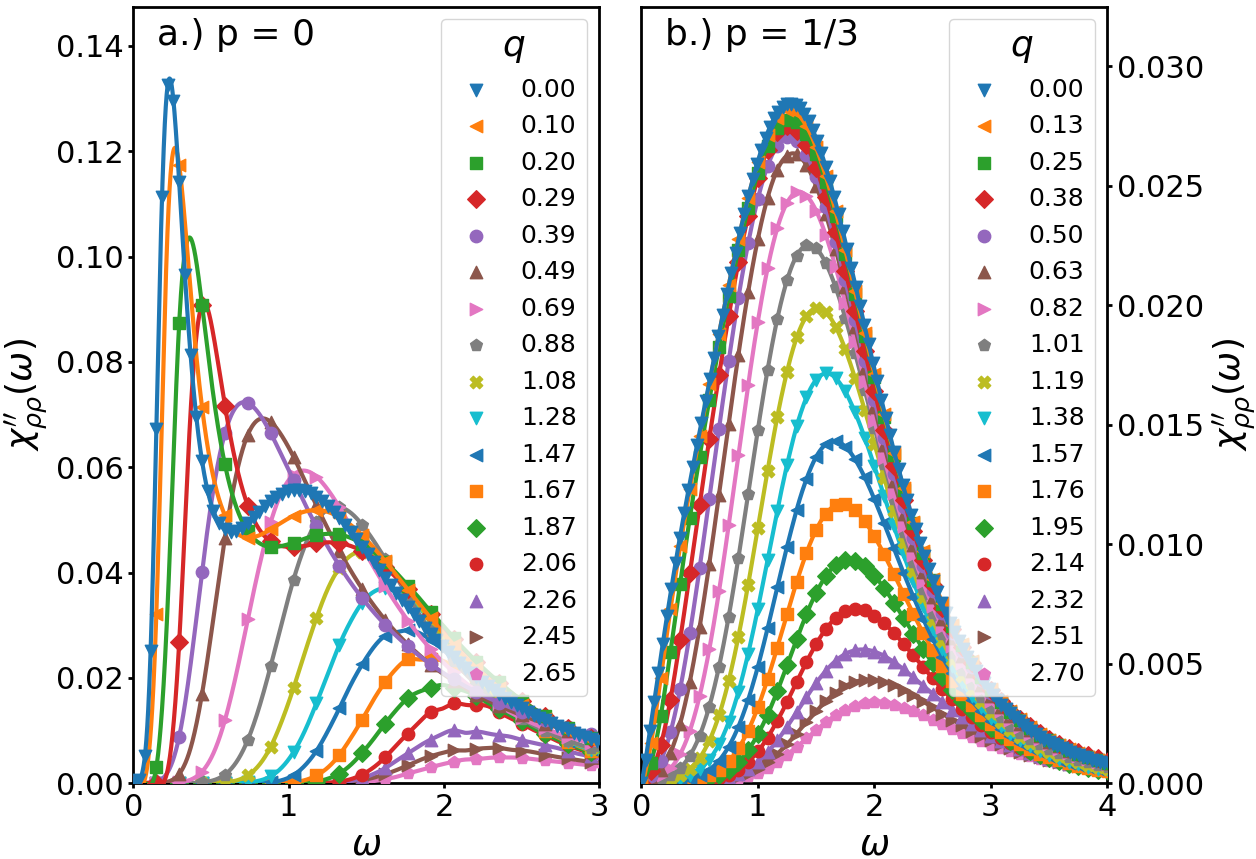}
\caption{Spectral densities for several values of the wave number $q$ for a.) $p=0$ and b.) $p=1/3$ at a fixed distance from criticality $r=-0.01$. Parameters analogous to those in Figures \ref{fig:clean_specs} and \ref{fig:dirty_specs}. A strong $q$-dependence can be seen in the clean case, whereas the diluted case features only weak $q$-dependence at the shortest wavelengths. These spectral densities are used as inputs for calculating the peak position dispersion $\omega_H(\mathbf{q})$.}
\label{fig:qvec-spectrals}
\end{figure}

\begin{figure}
\includegraphics[width=\linewidth]{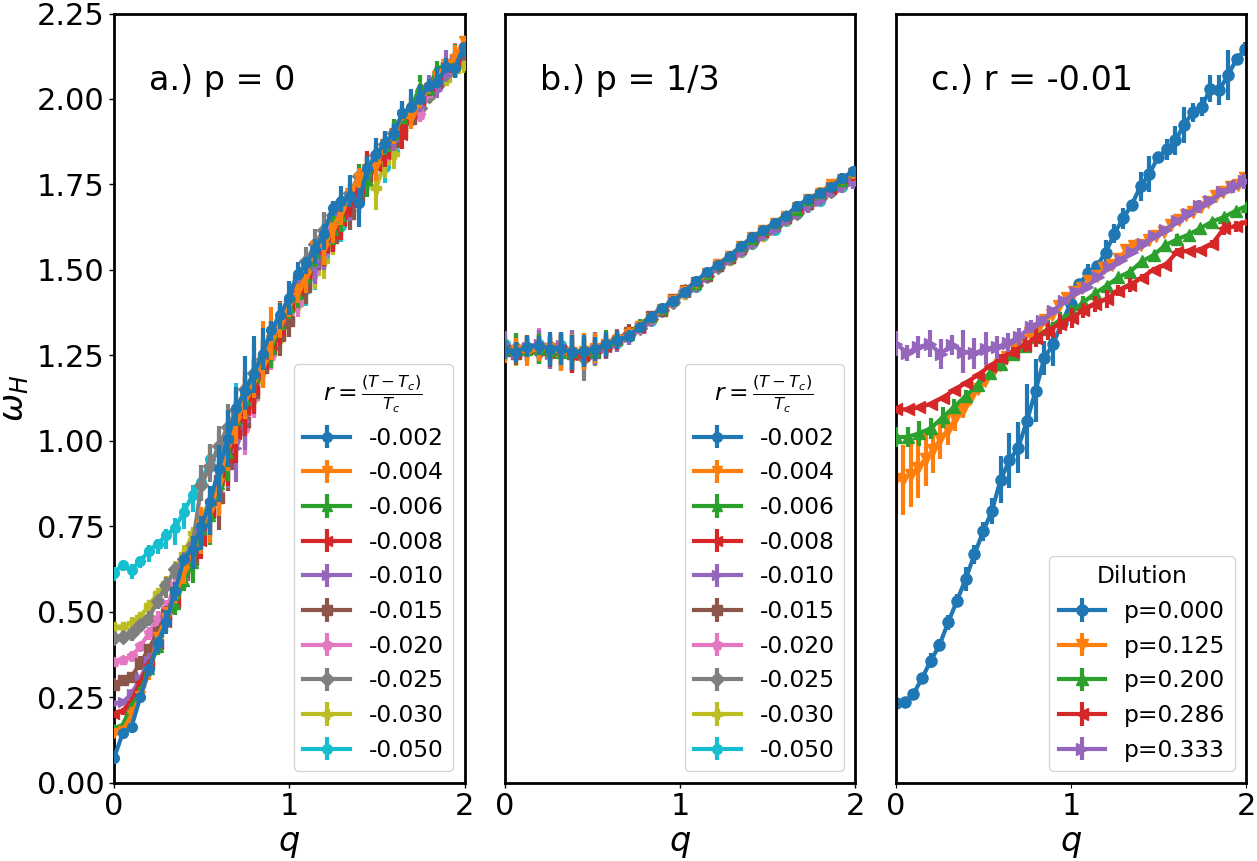}
\caption{a.) Clean case ($p=0$) dispersion of the Higgs energy $\omega_H$ at various distances from criticality. Calculated from a system of size $L=L_{\tau}=128$. b.) Diluted case ($p=1/3$) dispersion calculated from a system of size $L=100$, $L_{\tau}=452$. c.) Dispersion at a fixed distance from criticality ($r=-0.01$) for each of the dilutions considered. Error bars stem from variation of the fit parameter $\alpha$ as described in Sec.\ref{subsection:MaxEntError}. Parameters are as in Fig. \ref{fig:dirty_specs}.}
\label{fig:dispersion}
\end{figure}

To further understand the nature of this non-critical response we also study the dispersion $\omega_H(\mathbf{q})$ of the peak positions as a function of the wave vector $\mathbf{q}$. Figure \ref{fig:qvec-spectrals} shows spectral densities for $p=0$ and $p=1/3$ at a fixed distance from criticality $r=-0.01$ for several values of the wave number $q$. The clean case shows expected behavior for a $z=1$ quantum critical point. Namely, a quadratic long-wavelength dispersion $\omega_H(\mathbf{q}) = \omega_H(0) + a\mathbf{q}^2$ that crosses over to a linear form $\omega_H(\mathbf{q}) \sim |\mathbf{q}|$ as the critical point  $r \rightarrow 0$ is approached. The short-wavelength behavior is much more difficult to discern in our Monte Carlo data as peaks in the spectral densities have their amplitudes decreased and peaks broadened. Higgs mode dispersions of the clean system calculated within the critical scaling region can be seen in Figure \ref{fig:dispersion}, with error bars indicating estimated MaxEnt uncertainties. 

The diluted case exhibits different dispersion behavior with much weaker $\mathbf{q}$-dependence for short-wavelengths and showing nearly $\mathbf{q}$-independent behavior for low-energy, long-wavelength modes as illustrated in Figure \ref{fig:dispersion}b for $p=1/3$. The flattening of the dispersion below a critical $\mathbf{q}^*$ suggests the existence of a localization length $\lambda \sim 1/|\mathbf{q}^*|$ beyond which no Higgs excitations can extend. This behavior is contrasted with the clean case in Figure \ref{fig:dispersion}. Within the critical scaling region, the dispersions are essentially independent of $r$, further supporting the non-critical character of the Higgs excitations.

This localization behavior also shows dependence on the dilution strength $p$. Figure \ref{fig:dispersion}c shows dispersion relations for each of the dilutions considered, at a fixed distance from criticality $r=-0.01$. The effects of dilution are clearly drastic, as even the smallest dilution causes significant flattening of the dispersions for long wavelengths. The short-wavelength behavior is also interesting, as some cross-over effects may be significant. For the lowest dilutions, the flattening of the dispersions is still substantial for all wavelengths, but the long-wavelength response is still ``nearly-quadratic'' for $p=1/8$ with localization lengths only well-defined for $p=1/5$ and beyond. As dilution is increased, the localization length decreases monotonically as the dilution further inhibits long-range correlations. 

\section{Conclusion}
\label{section:Conclusion} 
We have conducted a study of the Higgs (amplitude) mode near the superfluid-Mott glass quantum phase transition in two-dimensions. To this end, we have considered a Bose-Hubbard model of disordered, interacting bosons in the limit of large integer filling. The resulting quantum rotor model is then mapped onto an equivalent $(2+1)$-dimensional classical XY model and simulated via standard Metropolis and Wolff Monte Carlo algorithms. Scalar correlation functions of the order parameter are calculated as a function of Matsubara frequency and the associated spectral densities are obtained via maximum entropy methods. 

In the clean case ($p = 0$), the spectral densities exhibit a sharp Higgs excitation. This excitation in the clean case satisfies scaling predictions near criticality and exhibits behavior in agreement with previous studies. Once disorder is introduced to the system, the spectral densities exhibit a broad, non-critical response that violates naive scaling arguments. This non-critical response is seen for all dilutions for which long range order is possible (i.e. below the lattice percolation threshold) and persists arbitrarily close to the critical point $r \rightarrow 0$, suggesting that the introduction of disorder to the system localizes the Higgs excitation. 

The possibility of disorder-induced localization of the Higgs mode is further supported by contrasting the dispersion of the maximum of the scalar susceptibility (the Higgs peak) as a function of wave vector for the clean and diluted cases. Expected behavior is observed in the clean case, with a quadratic dispersion for long wavelengths crossing over into a linear dispersion upon approaching the critical point. In contrast, dispersion in the diluted case shows a much weaker $q$-dependence and a nearly $q$-independent response for long wavelengths, implying a localization length below which no eigenmodes can extend. This localization broadens the spectral densities and prevents a critical response. This localization length is also observed to decrease monotonically with dilution strength, suggesting the Higgs mode becomes more strongly localized as the site-dilution further inhibits long-range correlations of the order parameter fluctuations. 

While the Monte Carlo results constitute strong evidence for the localization of the Higgs mode, further disentanglement of the source of the spectral response has been performed in a related work.\citep{PuschmannCrewse_etal, Puschmann2021} The work consists of the simulation of an inhomogeneous mean-field theory of the system (\ref{BH-hamiltonian}) that includes Gaussian fluctuations. The resulting spectral densities are analogous to the Monte Carlo results, showing a broad, non-critical response in the ordered phase arbitrarily close to the critical point. The mean-field theory permits the explicit analysis of the excitation eigenmodes which were found to be localized. Given that a mean-field theory has infinitely long-living excitations, this indicates localization as the source of the spectral density broadening. 

The effects of disorder on the Higgs mode has also been studied from a number of other theoretical and experimental perspectives. Swanson and collaborators\cite{Swanson_etal} have considered the fate of the Higgs mode across the disorder-induced superconductor-insulator transition by calculating complex conductivity $\text{Re}\sigma(\omega)$. In the clean case, the Higgs mode is predicted to give rise to an absorption threshold in the conductivity. This absorption threshold is not observed in the diluted case. Rather, excess spectral weight is observed for the sub-gap frequencies. The complex conductivity has also been studied experimentally in the disordered superconducting thin-films NbN and InO. This paper reports the observation of a critical Higgs mode after accounting for excess spectral weight in the complex conductivity arising from the superfluid condensate and quasiparticle dynamics. The experimental data were approximately reproduced in a Monte Carlo simulation of a Josephson junction Hamiltonian similar to (\ref{JJ-hamiltonian}). At first glance, the observation of the critical Higgs mode seems to contradict our results. However, the apparent observation of this Higgs mode is likely due to relatively weak disorder, with a maximum bond dilution of $p \approx 0.125$ considered in the simulations accompanying the experiment. For weak disorder, the system is expected to display a slow crossover from the clean to disorder behavior. Further study of this crossover region would be worthwhile, but requires considerably more computational effort. 

These results have the broader implication that disordered QPTs in general can exhibit unconventional collective excitations even in the case of conventional thermodynamic critical behavior. This motivates the further study of characteristics of this Higgs mode as well as the corresponding Goldstone modes. Additionally, it will certainly be interesting to investigate how spatial dimensionality and symmetries may affect these modes in disordered systems. Is it possible to classify disordered dynamics in a similar manner as the critical behavior? \cite{VojtaHoyos}

This work was supported in part by the NSF under grant No. DMR-1828489.

\bibliography{higgs-bib.bib}

\end{document}